\pdfoutput=1
\documentclass{PoS}

\title{Tuning improved anisotropic actions in lattice perturbation theory}

\ShortTitle{Tuning improved anisotropic actions}

\author{\speaker{Justin Foley} and {Colin Morningstar} \thanks{For the Hadron Spectrum Collaboration.}
\\
        Department of Physics, Carnegie Mellon University, Pittsburgh, PA 15213\\
        E-mail: \email{jfoley@andrew.cmu.edu}
				\\
        E-mail: \email{colin\_morningstar@cmu.edu}}

\abstract{We discuss the tuning of the anisotropic clover action 
					and Symanzik-improved gauge action in lattice perturbation theory.
					The fermion action is constructed from stout-smeared spatial links, 
					which complicates the calculation considerably. In addition, the 
					full quark-mass dependence of the action parameters is included 
					in this study. 
					We present results for the fermion and gauge aspect 
					ratios for varying bare aspect ratios, quark masses and 
					smearing parameters.
					}

\FullConference{The XXVI International Symposium on Lattice Field Theory\\
		 July 14-19 2008\\
		 Williamsburg, Virginia, USA}

\begin{document}

\section{Introduction}
In recent years, anisotropic lattice actions have seen increasing use in simulations.  
The $3+1$ anisotropic lattice, which has a temporal lattice spacing  $a_{t}$  that
is much less than the spatial lattice spacing $a_{s}$, is obviously very useful in studies 
of QCD at non-zero temperature, where it allows the temperature to be varied precisely at
fixed lattice spacing.  At zero temperature, the utility of 
anisotropic lattices in studies of the glueball spectrum of pure Yang-Mills is well-known~\cite{ColinandMike1,ColinandMike2}.
The fine temporal lattice spacing helps to resolve higher-lying states whose two-point 
functions quickly disappear into noise, while the relative coarseness 
of the spatial lattice spacing minimises the computational 
overhead. 

The ultimate goal of the Hadron Spectrum Collaboration is a precise determination 
of the low-lying spectrum of QCD. The low energy states of interest are not, 
however, restricted to just the ground state resonances in each irreducible 
representation of the lattice symmetry group, but include a number 
of higher-lying excitations. The use of anisotropic lattices with  fine temporal lattice spacings 
will be essential for the resolution of these states. 
It is also crucial that the action used should have a well-defined lattice transfer operator,
to avoid unphysical oscillations in the temporal fall-off of correlation functions. 
This requirement is potentially at odds with the Symanzik improvement program, which aims to eliminate 
lattice artifacts in a systematic way by adding irrelevant operators to the lattice action. 
However, on an anisotropic lattice, cutoff effects which depend on $a_{t}$ are highly suppressed
and the dominant lattice artifacts depend on the spatial lattice spacing only.
Therefore, a Symanzik-type improvement program can be implemented which removes these dominant cutoff effects,
without sacrificing positivity of the action.

The advantages associated with anisotropy come at a price. 
Anisotropic lattices break hypercubic symmetry, and accordingly 
the quark and gauge actions contain additional parameters which must 
be tuned so that the anisotropy (or aspect ratio) $ a_{s}/a_{t} $ 
measured using different physical probes takes a fixed target value. 
This tuning can be performed non-perturbatively~\cite{RobertandHueyWen}. However, in principle, 
non-perturbative tuning runs may be required for each new set of simulation parameters. 
In this proceedings, we describe the tuning of the anisotropic action parameters 
in one-loop perturbation theory. These lattice perturbative results are valid in the 
high $\beta$ regime. However, the ultimate goal of this work is to combine the results of lattice 
perturbation theory with the non-perturbative data to obtain functional forms for the 
action parameters which hold over much of parameter space.

\section{Actions}
The anisotropic quark action used in our simulations is 
\begin{eqnarray} 
	S_{\rm{quark}} &=& 
	a_{t} a_{s}^{3} \sum_{x} \bar{\psi} \left ( x \right )
	\left \{
	m_{0} + \gamma_{t} \nabla_{t} 
				- \frac{a_{t}} {2} \triangle_{t}
				+ \nu_{s} \sum_{k} \left ( \gamma_{k} \nabla_{k} - \frac{a_{s}} {2} 
				\triangle_{k} \right )  \right.
				\nonumber \\
				&& \hspace{2cm} \left. + 
				\frac{1} {2} \left [  
				c_{t} a_{s} \sum_{k} \sigma_{t k} F_{t k} 
				+ 
				c_{s} a_{s} \sum_{k<l} \sigma_{k l} F_{k l}
				\right ] \right \} \psi \left ( x \right ),
\end{eqnarray}
where the covariant derivatives and clover-leaf discretisation of the field strength 
tensor are built from link variables which have been stout-smeared~\cite{ColinandMike3} in the spatial 
directions. 
As noted above, it is important that the temporal links be left unsmeared.
Furthermore, the action includes tadpole improvement factors for the spatial links, 
although, 
for an action constructed from smeared links, these are close to unity.

In a numerical simulation all lengths are expressed in lattice units, and 
the quark action depends on a bare anisotropy parameter which we denote 
$ \xi_{0} $. This bare parameter can receive radiative corrections.
However, it is possible to choose the coefficient $ \nu_{s} $, multiplying the 
kinetic term, such that the aspect ratio measured from the quark dispersion 
relation takes a predefined target value.
At tree level and in the chiral limit, 
setting $ \nu_{s} = 1$ fixes the measured aspect ratio to the bare anisotropy. 
The `clover' coefficients $ c_{t} $ and $ c_{s} $ are tuned such that 
on-shell quantities are free of $ \mathcal{O} \left ( a_{t}, a_{s} \right ) $ 
cutoff effects. For massless quarks, their tree-level values are 
$ c_{t} = \frac{1} {2} \left ( \nu_{s} + \frac{1} {\xi_{0}} \right )$, 
$ c_{s} = \nu_{s} $.

The gauge action incorporates Symanzik and tadpole improvement, and can be 
written
\begin{eqnarray} 
	S_{\rm{gauge}} 
	&=& 
	- \beta
	\left \{
	\xi_{0} \left [   
	\frac{4} {3} \sum_{i} P_{t i}  - \frac{1} {12} \sum_{i} R_{t i}
	\right ]
	+ 
	\frac{1} {\xi_{0}} 
	\left [
	\frac{5} {3} \sum_{i < j} P_{i j} 
	- \frac{1} {12} \sum_{i < j} \left( R_{i j} + R_{j i} \right )
	\right ]
	\right \}.
\end{eqnarray}
$ R_{i j} $ denotes a $ 1 \times 2 $ rectangle, two links long in 
the $ j $ direction, summed over all lattice sites. 
The dominant discretisation errors of this action appear at 
$ \mathcal{O} \left ( a_{t}^{2}, a_{s}^{4}, \alpha_{s} a_{s}^{2} 
\right ) $. Crucially, the lagrangian is just one link wide
in the temporal direction, which guarantees a well-defined 
single-timeslice transfer operator. 

\section{Tuning the quark action parameters}
In lattice perturbative studies, for the sake of simplicity, one 
often ignores the  dependence of the action parameters on the bare quark mass.
In the Fermilab formalism~\cite{ElKhadra:1996mp}, on the other hand, the full quark-mass dependence 
of the action parameters is determined, yielding an action which can be used 
in both the chiral and heavy-quark regimes. 
In our simulations, $ m_{0} a_{s} $ is not always guaranteed to be very small, 
and we have adopted a Fermilab-type approach in this study.
Even at the tree-level, the parameters of the quark action receive non-trivial 
mass-dependent corrections. 
The tree-level mass dependence of the kinetic coefficient $ \nu_{s} $ can be determined 
by expanding the free-quark energy in powers of the spatial momentum, and demanding that the 
quark rest mass and kinetic mass be equal for a given aspect ratio:
\begin{eqnarray}
	a_{t} E \left ( \vec{p} \right ) = a_{t} M_{\rm{rest}} 
	+ \frac{1} {2 a_{t} M_{\rm{kin}}}  
	\frac{| a_{s} \vec{p} |^{2}} {\xi_{0}^{2}} + \cdots.
\end{eqnarray}

This implies that
\begin{eqnarray} 
	\nu_{s}^{(0)} =
	\sqrt{ \frac{1} {4} \xi_{0}^{2} \sinh^{2} \left ( a_{t} M_{\rm{rest}}^{(0)} \right )  
		+ \exp \left( a_{t} M_{\rm{rest}}^{(0)} \right ) 
		\frac{
		\sinh \left ( a_{t} M_{\rm{rest}}^{(0)} \right ) }
		{ a_{t} M_{\rm{rest}}^{(0)} }
	}
		- 
		\frac{\xi_{0}} {2} \sinh \left ( a_{t} M_{\rm{rest}}^{(0)} \right ),
	\label{eqn:numass}
\end{eqnarray}
where $ a_{t} M_{\rm{rest}}^{(0)} = \log \left( 1 + a_{t} m_{0} \right ) $.
Expanding this expression, one finds that the leading order correction to 
$ \nu_{s}^{(0)} $ is linear in the bare mass.
The tree-level chromoelectric and chromomagnetic coefficients can be determined 
by requiring that, at low momenta, scattering amplitudes match their continuum counterparts. 
They are
\begin{eqnarray}
	c_{t}^{(0)} &=& \frac{ a_{t} m_{0} \left ( 2 + a_{t} m_{0} \right ) } 
	{ 4 \nu_{s}^{(0)} \xi_{0} \left [  \log \left (1 + a_{t} m_{0} \right ) \right ]^{2} }
	- 
	\frac{ \nu_{s}^{(0)} } { \xi_{0} a_{t} m_{0} \left ( 2 + a_{t} m_{0} \right ) }, 
	\nonumber \\
	c_{s}^{(0)}  &=& \nu_{s}^{(0)}.
	\label{eqn:cmass}
\end{eqnarray}

\begin{figure}
\begin{center}
\includegraphics[width=10cm]{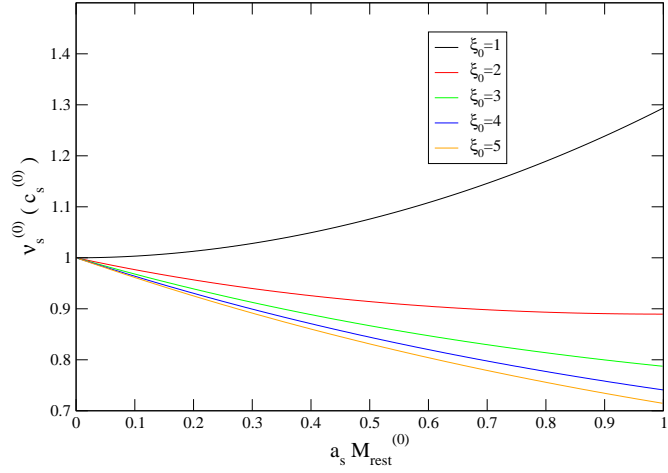}
\caption{$\nu_{s}^{(0)}$ as a function of the tree-level quark rest mass 
given in spatial lattice units. \label{fig:treelevelnu}}
\end{center}
\end{figure}

At tree level, our quark action is essentially a reparametrisation of the actions 
used in Refs.~\cite{ElKhadra:1996mp,kronfeld}, and accordingly Eq.~\ref{eqn:numass} and Eq.~\ref{eqn:cmass}  are in agreement with the 
results of those papers. Fig.~\ref{fig:treelevelnu} plots $ \nu_{s}^{(0)} $ versus the tree-level rest mass for a 
number of bare anisotropy values. Note the quark mass dependence of $\nu_{s}^{(0)}$ even in the isotropic limit. 

To go beyond tree level, we compute the quark energy and scattering 
amplitudes in perturbation theory and apply the same tuning criteria. 
Our calculation of the leading order correction to $ \nu_{s} $ therefore
amounts to a determination of the quark self-energy in one-loop  
perturbation theory.

\section{Gauge anisotropy} 
The anisotropy parameter for the gauge action used in this study 
has previously been determined to one-loop order in pure Yang-Mills in Ref.~\cite{Ron}. 
In that study, twisted boundary conditions~\cite{LW} were used as an infrared regulator for the 
gluons. The anisotropy was determined by demanding that one of the  
stable states in the twisted 
world, the so-called A meson, satisfy a relativistic dispersion relation. Determining the 
anisotropy, therefore amounts to a calculation of the one-loop gluon self-energy. 
We have repeated that calculation and extended it to include quark-loop effects.

\section{Methodology} 
Lattice perturbative calculations are notoriously complicated, even at the 
one-loop level. In our case, although only a moderate number of Feynman diagrams arise, the 
vertex functions which appear in these diagrams are extremely complicated due to smearing of the 
link variables in the quark action and Symanzik improvement of the gauge action. 
For such calculations, the best approach is to automate the derivation of vertex functions 
and the evaluation of the final momentum integrals as much as possible~\cite{LW,Colin,Ron2}.

To evaluate the vertex functions, we employed a number of independent methods. 
In one approach, we used the suite of Python code described in Ref.~\cite{Ron2} to expand the 
actions to the required order in the coupling. We have written a parser which converts
the resulting data file into a method of a corresponding C++ vertex class. 
As a cross-check,
a completely separate suite of C++ code has been developed
which evaluates the momentum space vertex functions for a given set of four-momenta. 
Using this automated approach, we can evaluate vertex functions for an arbitrary level 
of link smearing. In fact, we have checked that it is possible to evaluate a four-gluon vertex function 
for the quark action
with one hundred iterations of the stout-link smearing algorithm in just a few seconds.

Spin trace evaluations are completely automated, and the results 
presented here have been obtained using the Vegas integration routine~\cite{Lepage}. 
Where derivatives of the self-energy with respect to external momentum 
are required, automatic differentiation can be applied directly to the integrand 
being passed to Vegas. 
However, the resulting sharply peaked function 
can prove difficult to integrate. In that case, more precise results 
may be obtained by evaluating the self-energy at different values 
of the external momentum and estimating the derivative numerically~\cite{Us}.

All calculations are performed in a Lorentz-covariant gauge and, where 
practicable, we repeat calculations in both Feynman and Landau gauge 
to verify the gauge-invariance of our results.
\section{Results}

\begin{figure} 
\begin{center}
	\includegraphics[width=11.8cm]{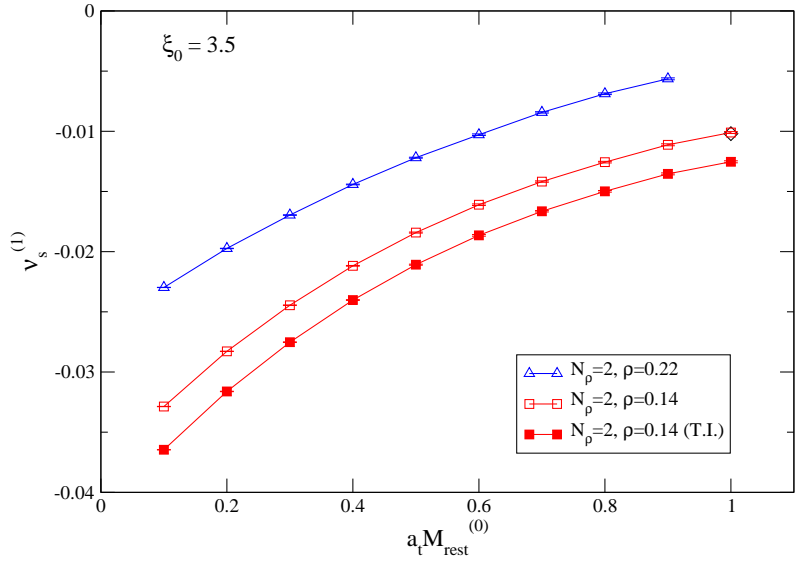} 
	\label{fig:nu1}
	\caption{One-loop correction to $ \nu_{s} $ as a function of quark mass. T.I. denotes tadpole 
						improvement, while the hollow symbols denote results obtained before tadpole improvement.
						\label{fig:nu1}}
\end{center}
\end{figure}

\begin{figure} 
	\begin{center}
		\includegraphics[width=11.8cm]{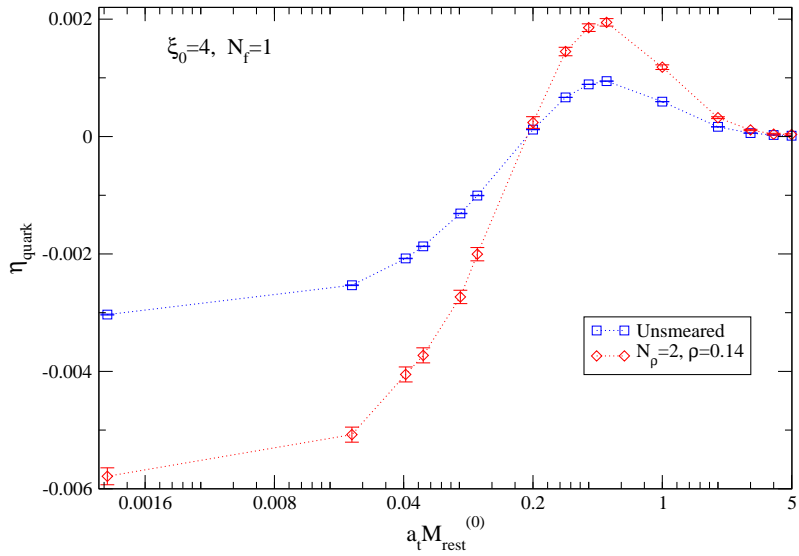}
		\label{fig:vary_mass}
		\caption{The quark-loop contribution to $ \eta $
		plotted as a function of the tree-level quark rest mass for different levels of smearing. 
		As described in the text, $ \eta $ is defined by $ \xi_{g}/\xi_{0} = 1 + g^{2} \eta $, where $\xi_{g}$ denotes the anisotropy 
		measured from the gluonic dispersion relation.
			\label{fig:vary_mass}
		}
	\end{center}
\end{figure}

A plot of the one-loop correction to $\nu_{s}$ as a function of the quark 
rest mass for a typical anisotropy value is shown in Fig.~\ref{fig:nu1}.
The data presented here are coefficients of $ g^{2} $.
Two different levels of smearing are shown. The stout link smearing 
parameters of $ n_{\rho}=2 $ and $ \rho = 0.14 $ were found to 
simultaneously minimise the additive quark mass renormalisation and 
maximise the spatial plaquette in one-loop perturbation theory~\cite{RobertandHueyWen}. 
For this choice of smearing parameters, the plot shows that tadpole 
improvement has little effect on $ \nu_{s} $. In this study, as in our 
simulations, we define the tadpole factor to be the fourth root of the 
expectation value of the plaquette.

Regarding corrections to the gauge anisotropy parameter, there are a couple of 
points which are important to note.
First, the pure gauge contribution to the anisotropy and the contribution 
from sea quarks are additive.
That is to say, the correction for full 
QCD is simply the sum of the correction coming from pure Yang-Mills and 
the quark-loop contribution. Second, it is obvious that the 
magnitude of the correction coming from quark loops is proportional 
to $N_{f} $, so that it increases with an increasing number of flavours. 
It is also important to note that, at one-loop order, the sea-quark 
contribution to the gauge anisotropy is independent of the choice 
of gauge action. Therefore, the quark-loop corrections presented here hold for 
any choice of anisotropic gauge action.

Following Ref.~\cite{Ron}, we define $ \eta $ to be the one-loop correction to the anisotropy divided 
by the bare anisotropy value appearing in the gauge action. 
As a check on our methods, we calculated the pure gauge contribution 
to $ \eta $ 
as described in that reference and found agreement with the
results given there.
The contribution of a single sea-quark flavour to this quantity for a fixed bare anisotropy but a 
varying sea-quark mass is shown in Fig.~\ref{fig:vary_mass}. As expected, this contribution goes to zero 
in the heavy quark limit, but becomes significant at light 
quark masses. 
At sufficiently light quark masses the contribution to $ \eta $ from three degenerate quark flavours can match the purely 
gluonic contribution in magnitude~\cite{Us}.

\section{Conclusion and outlook} 
Anisotropic lattice actions must be tuned to guarantee the restoration of 
Euclidean invariance in the continuum limit.
We have described this tuning in one-loop perturbation 
theory. To remain as close to simulation as possible, the full quark 
mass dependence has been included in this calculation.
Moreover, we have developed a suite of software capable of handling complicated 
actions involving an arbitrary level of link smearing.
One-loop calculations of the quark action improvement coefficients 
$c_{s}$ and $c_{t}$ using this machinery are currently in progress.

Future work will obviously include a comparison with  non-perturbative results. 
It will also be useful to know how the addition of an adjoint term 
to the gauge action, as outlined in Ref.~\cite{ColinandMike4}, affects the tuning of 
the gauge anisotropy.

\section{Acknowledgments} 
We are grateful to the authors of Ref.~\cite{Ron2} for providing us with 
a copy of the `HiPPy' Python code, which was used in the derivation of the vertex functions.
This work was supported by the U.S. National Science Foundation under the Awards PHY-0510020 
and PHY-0653315.

\end{document}